\documentclass[aps,pre,onecolumn,10pt]{revtex4-1}
\usepackage{amsmath}
\usepackage{graphicx}
\usepackage{color}
\usepackage{hyperref}
\usepackage{psfrag}
\usepackage{stmaryrd}
\usepackage{amssymb}
\usepackage{wasysym}
\usepackage{multirow}

\def \mps {m~s$^{-1}$}

\usepackage{geometry}
\begin{document}

\title{Viscosity effects in wind wave generation}
\author{A. Paquier}
\author{F. Moisy}
\author{M. Rabaud}
\affiliation{Laboratoire FAST, CNRS, Universit\'e Paris-Sud, Universit\'e Paris-Saclay, 91405 Orsay, France.}

\date{\today}

\begin{abstract}

We investigate experimentally the influence of the liquid viscosity on the problem of the generation of waves by a turbulent wind at the surface of a liquid, extending the results of Paquier, Moisy and Rabaud [Phys. Fluids {\bf 27}, 122103 (2015)] over nearly three decades of viscosity. The surface deformations are measured with micrometer accuracy using the Free-Surface Synthetic Schlieren method.  We recover the two regimes of surface deformations previously identified: the wrinkles regime at small wind velocity, resulting from the viscous imprint on the liquid surface of the turbulent fluctuations in the boundary layer, and the regular wave regime at large wind velocity. Below the wave threshold, we find that the characteristic amplitude of the wrinkles scales as $\nu^{-1/2} u^{* 3/2}$ over nearly the whole range of viscosities, whereas their size are essentially unchanged. We propose a simple model for this scaling, which compares well with the data.  We finally show that the critical friction velocity $u^*$ for the onset of regular waves slowly increases with viscosity as $\nu^{0.2}$. Whereas the transition between wrinkles and waves is smooth at small viscosity, including for water, it becomes rather abrupt at large viscosity. Finally, a new regime is found at $\nu > 100-200 \times 10^{-6}$~m$^2$~s$^{-1}$, characterized by a slow, nearly periodic emission of large-amplitude isolated fluid bumps.

\end{abstract}

\maketitle

\section{Introduction} \label{sec:intro}

When wind blows over a liquid surface, waves can be generated and propagate downstream. Despite the fact that this simple phenomenon has inspired over a century of research, understanding the physics of wind wave generation is still a standing challenge \cite{Sullivan_2010}. A key issue is the determination of the critical velocity $U_c$ above which waves are generated, and the influence of the liquid viscosity on this critical velocity.  It is well known that the inviscid prediction by the simple Kelvin-Helmholtz model~\cite{Helmholtz_1868,Kelvin_1871}, $U_c \simeq 6.6$~\mps~ for the air-water interface,  largely overestimates the actual threshold observed at sea or in laboratories, typically in the range $1-3$~\mps. However, including viscous corrections to this model does not solve the discrepancy: it only leads to marginal modification of the critical velocity~\cite{Lindsay_1984, Funada_2001, Kim_2011}, in clear contradiction with experimental observations of higher wind velocity thresholds for more viscous fluids~\cite{Keulegan_1951,Francis_1956,Gottifredi_1970,Kahma_1988}. A clear understanding of the influence of the fluid viscosity on the critical velocity is still lacking.
 
Another open question is the nature of the surface deformations below the wave generation threshold. It would be erroneous to assume that the surface remains entirely flat at low wind velocity: slight surface deformations are often observed in experiments~\cite{Keulegan_1951, Kunishi_1963, Hidy_1966, Plate_1969, Gottifredi_1970, Wu_1978, Kahma_1988, Ricci_1992_PhD, Caulliez_1998, Lorenz_2005}, breaking the perfect mirror reflection over a water surface even below the wave onset. Indeed, even for moderate wind velocity, the flow in the air is generally turbulent, and these small surface deformations, which we call {\it wrinkles}, are the imprints of the pressure and shear stress fluctuations at the surface. Such wrinkles are also found at short time, before the onset of waves, in direct numerical simulations of temporally growing waves~\cite{Lin_2008,Zonta_2015}. This noisy state below the wave onset may explain the substantial scatter in the velocity threshold reported in the literature.

In spite of their ubiquity in experiments, the properties of these wrinkles below the wave onset have not been quantified so far, because of their very small amplitude (typically 1-10$~\mu$m) well below the resolution of classical one-point measurement devices, and their high sensitivity to mechanical vibrations of the experimental setup~\cite{Kahma_1988, Caulliez_1998}.  The first quantitative analysis of these wrinkles below the wave onset was provided by Paquier {\it et al.}~\cite{Paquier_2015}, in the case of a liquid thirty times more viscous than water. Taking advantage of the excellent vertical resolution of the optical method Free-Surface Synthetic Schlieren (FS-SS)~\cite{Moisy_2009}, a spatio-temporal analysis of these wrinkles was performed, and the transition between these wrinkles and the well-defined waves with crest perpendicular to the wind direction was characterized in detail.  The amplitude of the wrinkles was found to be essentially independent of the fetch (the distance upon which the air blows on the liquid), and to increase approximately linearly with wind velocity. Although we naturally expect a weaker imprint of the applied stress fluctuations on the surface of a more viscous liquid, the scaling of the wrinkles amplitude with viscosity is not known to date.

In this paper, we explore systematically the influence of the liquid viscosity on the main properties of the surface deformations generated by a turbulent boundary layer in the air, both below and above the wave generation threshold. The aim is to extend over a wide range of viscosities ($\nu = 0.9 - 560 \times 10^{-6}$~m$^2$~s$^{-1}$) the results of Paquier {\it et al.}~\cite{Paquier_2015} obtained for a single viscosity ($\nu = 30 \times 10^{-6}$~m$^2$~s$^{-1}$), in order to gain insight into the physical mechanisms governing the dynamics of the wrinkles and the transition to the regular wave regime. We observe that the characteristic amplitude of the wrinkles scales as $\nu^{-1/2} u^{*3/2}$, where $u^*$ is the friction velocity, and that the critical velocity for wave generation slowly increases as $\nu^{0.2}$ over almost all the range of viscosities, from $\nu = 10^{-6}$~m$^2$~s$^{-1}$ (water) to approximately $200 \times 10^{-6}$~m$^2$~s$^{-1}$. At larger viscosity, the nature of the wave transition is found to evolve towards a new regime, characterized by a slow, nearly periodic emission of large-amplitude isolated fluid bumps.

\section{Experiments} \label{sec:exp}

\begin{figure}[t]
	\begin{center}
		\includegraphics[width=\columnwidth]{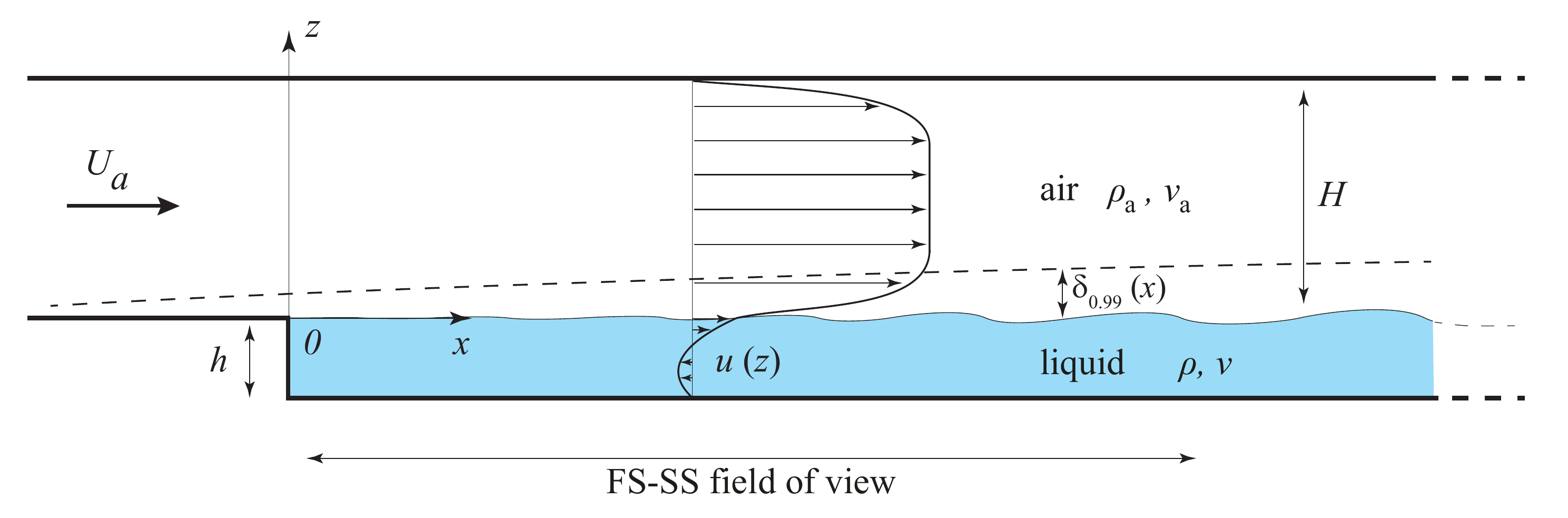}
		\caption{Sketch of the system. The velocity profiles in the air and in the liquid are not drawn to scale: the air velocity is typically 100 times larger than the liquid velocity. The wave amplitude is exaggerated.}
		\label{fig:system}
	\end{center}
\end{figure}

The experimental set-up, sketched in Fig.~\ref{fig:system}, is the same as in Paquier {\it et al.}~\cite{Paquier_2015,Paquier_2016_PhD} and is only briefly described here. It is composed of a fully transparent Plexiglas rectangular tank, fitted to the bottom of a horizontal channel of rectangular cross-section. 
The tank is of length $L=1.5$~m, width $W=296$~mm, and depth $h=35$~mm and the channel height is $H=105$~mm, with its width identical to that of the tank. The tank is filled with liquid mixtures of different viscosities, such that the surface of the liquid precisely coincides with the bottom of the wind tunnel.
Air is injected upstream at a mean velocity $U_a$ that can be adjusted in the range $1-10$~\mps. 
We define the coordinate axes $(x, y, z)$ in the streamwise, spanwise and upwards vertical direction, respectively. The origin (0,0,0) is located at the free surface at $x=0$ at mid-distance between the lateral walls of the channel.  In accordance with the literature, the term \textit{fetch} refers to the distance over which the wind blows (in our case, the distance $x$ from the beginning of the tank).

The surface deformation of the liquid is measured using the FS-SS method~\cite{Moisy_2009}. A random dot pattern is located under the tank and is imaged through the interface by a fast camera. Comparing the refracted image of the pattern to a reference image taken when the interface is flat (in absence of wind), the FS-SS allows to obtain the height field $\zeta(x,y,t)$.  The acquisitions are performed over a field of view elongated in the $x$ direction, of dimensions $390 \times 280$~mm, with horizontal resolution of 3~mm. The resolution in $z$, controlled by the surface-pattern distance (chosen between 6 and 29~cm), is of order of 1~$\mu$m for the smaller wave amplitudes, in the wrinkle regime.

In order to investigate the influence of viscosity on wind wave generation, the kinematic viscosity $\nu$ of the liquid is changed from $0.9$ to $560 \times 10^{-6}$~m$^2$~s$^{-1}$, using differently concentrated mixtures of glycerol and water or Glucor 60/80HM (a glucose syrup referred later simply as glucor) and water. Glycerol-water mixtures are used for viscosity up to about $120 \times 10^{-6}$~m$^2$~s$^{-1}$ and glucor-water mixtures for higher viscosity. Indeed, due to the hygroscopic nature of glycerol~\cite{SDA_1990} and the rapid evolution of the viscosity of a glycerol-water solution with concentration and temperature~\cite{Segur_1951}, we had to switch to glucor, much more stable in time at high concentration in water.

Contrary to their kinematic viscosity, the density of the different mixtures does not change much, from 1.0 to $1.36 \times 10^3$ kg~m$^{-3}$. The kinematic viscosity of the glycerol-water mixtures are taken from the tables for the actual values of $\rho$ and $T$ while the kinematic viscosity of the glucor-water mixtures is measured by a rheometer Anton Paar Physica MCR 501. The temperature is controlled over the duration of an experiment to $\pm 0.5^{\circ}$C. More details on the parameters of each experiment are given in Ref.~\cite{Paquier_2016_PhD}.

In the wide range of liquid viscosities covered here, the wave behavior ranges from essentially inviscid to strongly damped. According to Leblond and Mainardi~\cite{Leblond_1987}, the attenuation length of a typical wavelength $\lambda \simeq 30$~mm is about 2~m for water (i.e., the waves are essentially undamped over the size of the tank) and decreases down to 8~mm for the most viscous liquid used here. On the other hand, the influence of viscosity on the frequency (and hence on the phase velocity) is less pronounced: for this typical wavelength the frequency is essentially given by the inviscid prediction up to $\nu \simeq 10^{-4}$~m$^2$~s$^{-1}$, and is decreased by 40\% for the most viscous liquid used here. The cut-off wavelength, below which waves are over-damped and cannot propagate, is significantly smaller than the typical wavelengths observed over much of the viscosity range, and is not expected to influence the results; this cut-off becomes however significant ($\sim 12$~mm) for the most viscous liquid, so its influence on the wave propagation may be visible.

An important aspect of this setup is the presence of a steady recirculation flow in the liquid induced by the mean wind shear stress at the surface. At sufficiently large liquid viscosity, this flow is essentially laminar, and is well described by the Couette-Poiseuille solution far from the side and end walls of the tank: the mean velocity profile, measured for $\nu = 30 \times 10^{-6}$~m$^2$~s$^{-1}$ in Ref.~\cite{Paquier_2015}, follows the expected parabolic law, $u(x,z)=U_s(x) (1+z/h) (1+3z/h)$ for $-h \leq z \leq 0$, where $U_s(x) = u(x,z=0)$ is the surface drift velocity resulting from the wind shear stress. This velocity profile is nearly homogeneous in $x$ and $y$ except at very small fetch (on a distance of the order of the liquid height) and over the last 30~cm of the tank (where surface contamination cannot be avoided). The surface drift $U_s$ remains small in our experiments: for $U_a \simeq 3$~\mps, in the wrinkle regime, $U_s$ is smaller than 5~cm~s$^{-1}$ for $\nu > 3 \times 10^{-6}$~m$^2$~s$^{-1}$, resulting in a Reynolds number $Re_s = U_s h/\nu$ smaller than 1000, for which the Couette-Poiseuille flow is expected to remain stable~\cite{Klotz_2016}. This stability is not guaranteed however at lower viscosity, in particular in the case of water, for which $Re_s$ reaches $5000$.

Finally, we assume that the velocity profile in the air, characterized in detail for $\nu = 30 \times 10^{-6}$~m$^2$~s$^{-1}$ in Ref.~\cite{Paquier_2015}, is not significantly affected by the change of liquid viscosity. The boundary layer, which is tripped using sandpaper located at $x = - 260$~mm, is already turbulent when it reaches the liquid at $x=0$. It remains similar to that of a classical turbulent boundary layer developing over a no-slip flat wall, at least in the wrinkle regime, because the surface drift velocity $U_s$ is comfortably smaller than $U_a$, and also because the amplitude of the wrinkles remains much smaller than the thickness of the viscous sublayer ($\delta_v = \nu_a / u^* \simeq 0.05- 0.3$~mm, with $\nu_a$ the kinematic viscosity of the air and $u^*$ the friction velocity, discussed in Sec.~\ref{sec:wr}). We can therefore assume that the dependence on viscosity of the shape and amplitude of the surface deformations, at least in the wrinkle regime, is essentially governed by the liquid response to an otherwise (statistically) identical turbulent air flow.

\section{Experimental results}\label{sec:exp_results}

\subsection{The three wave regimes}

\begin{figure}[htbp]
	\begin{center}
		\includegraphics[width=0.92\columnwidth]{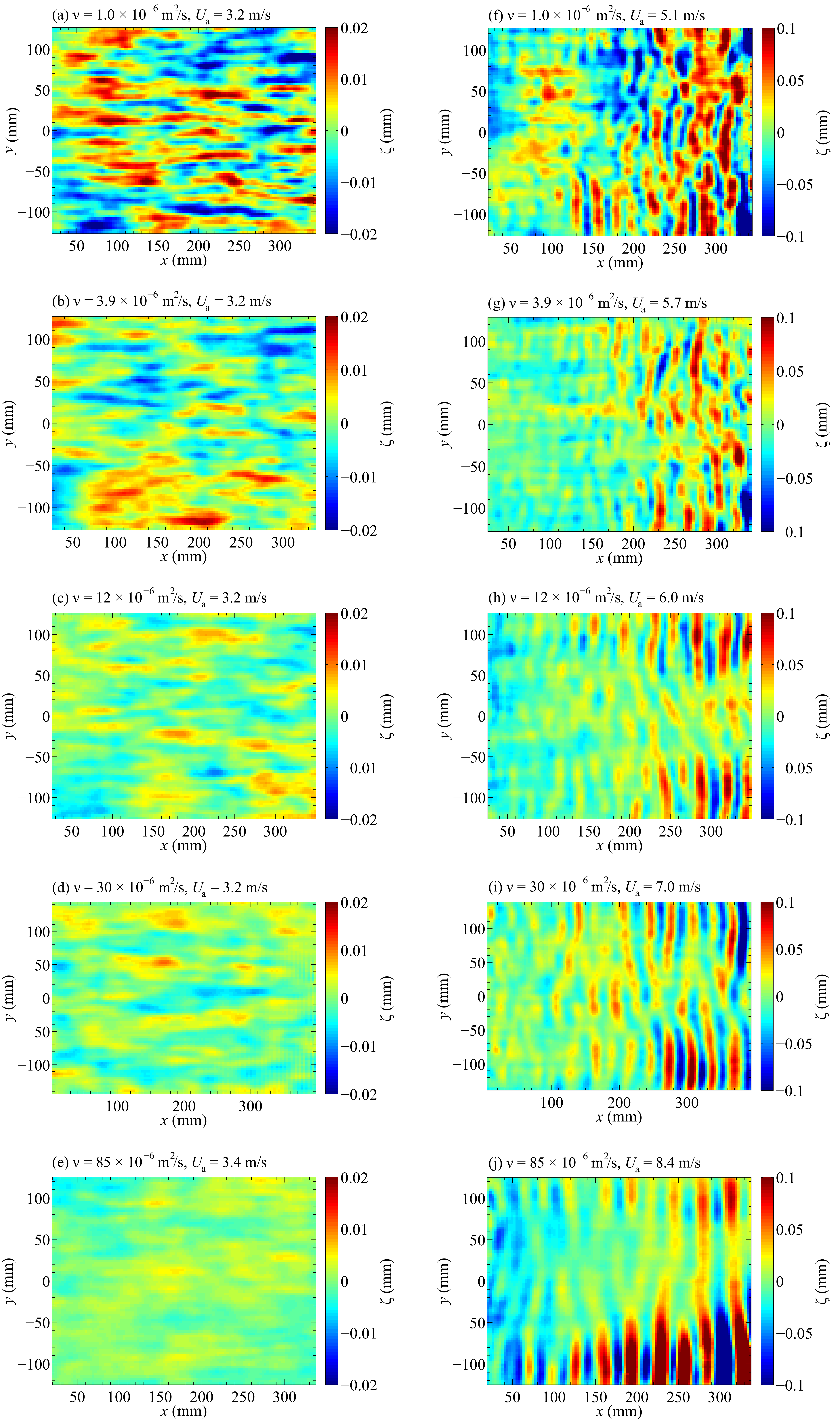}
		\caption{Instantaneous surface height $\zeta(x,y)$ measured by FS-SS, for increasing fluid viscosities ($\nu = $~1.0, 3.9, 12, 30 and 85~$\times 10^{-6}$~m$^2$~s$^{-1}$, from top to bottom); (a-e) below the threshold, at a wind velocity chosen to be about 3.2~\mps ($u^* \simeq 0.16$~\mps), showing disorganized wrinkles elongated in the streamwise direction; (f-j) slightly above the (viscosity-dependent) threshold velocity, showing spatially growing regular waves. Note that the color map is the same on a column but varies between the left and right column.}
		\label{fig:FSSS_different_visc}
	\end{center}
\end{figure}

Figure~\ref{fig:FSSS_different_visc} shows snapshots of the surface deformation for increasing liquid viscosities, from $\nu = 10^{-6}$ up to $85 \times 10^{-6}$~m$^2$~s$^{-1}$, below and just above the (viscosity-dependent) wave onset velocity. In this range of viscosities, the surface deformation fields are qualitatively similar to that reported in Ref.~\cite{Paquier_2015} for $\nu = 30 \times 10^{-6}$~m$^2$~s$^{-1}$: at low wind velocity, below the wave onset (left column,  Fig.~\ref{fig:FSSS_different_visc} (a-e)),  the surface field is composed of disorganized wrinkles, elongated in the streamwise direction, of amplitude essentially independent of $x$.  As expected, the wrinkles amplitude decreases as the liquid viscosity is increased, typically from 20 to 2~$\mu$m. At larger wind velocity, above the wave onset (right column, Fig.~\ref{fig:FSSS_different_visc} (f-j)), the surface field shows spatially growing waves with crests nearly perpendicular to the wind direction. At small viscosity, in particular for water, these regular waves coexist with the wrinkles, as illustrated in Fig.~\ref{fig:FSSS_different_visc}(f) at small fetch, resulting in a mixed wave pattern near the onset. As the viscosity is increased, the amplitude of the wrinkles decreases and the transition to the regular wave regime becomes more visible.

\begin{figure}[bt]
	\begin{center}
		\includegraphics[width=0.6\columnwidth]{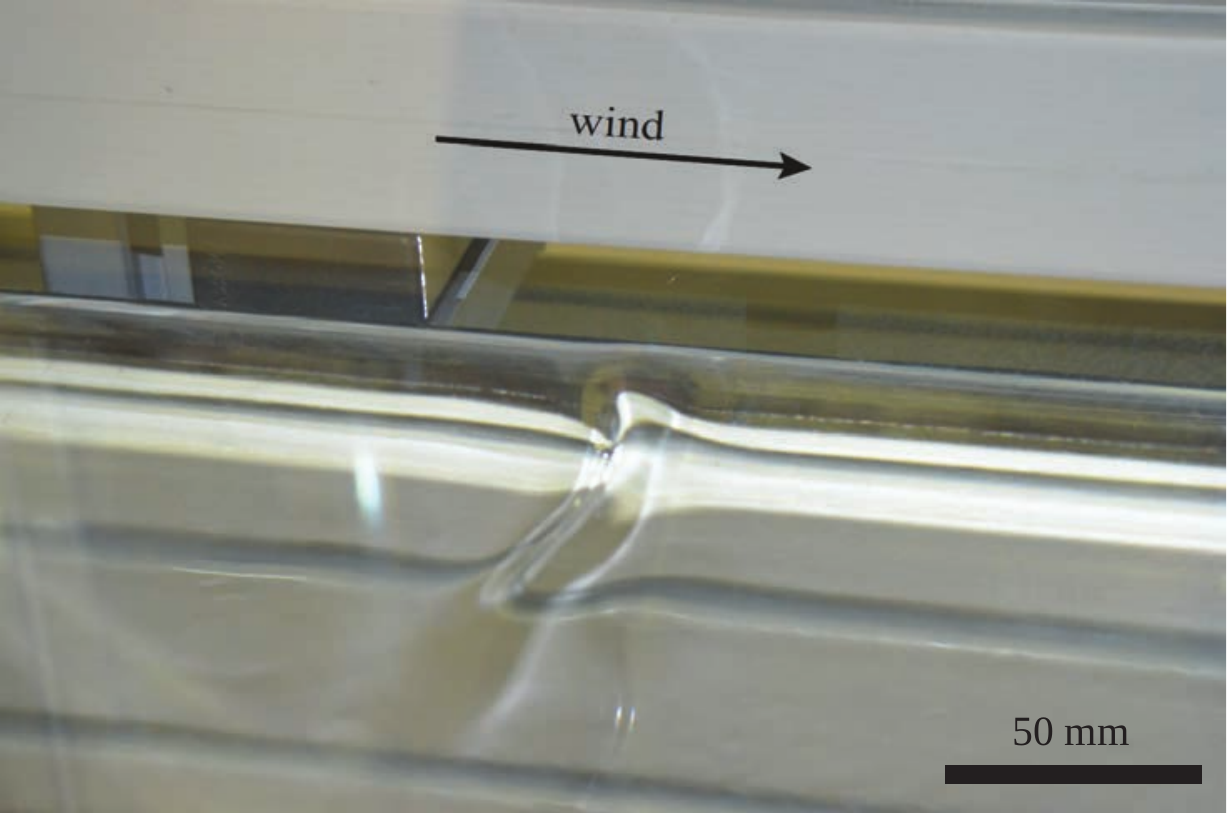}
		\caption{Picture of a solitary wave in the highly viscous regime ($\nu=400 \times 10^{-6}$~m$^2$~s$^{-1}$, $U_a=10.6$~m~s$^{-1}$).}
		\label{fig:solitarywave}
	\end{center}
\end{figure}

This general picture holds for fluid viscosity up to $\nu \simeq 100-200 \times 10^{-6}$~m$^2$~s$^{-1}$. At larger viscosity, the wrinkle regime is qualitatively similar, but the transition to the regular wave regime as the wind velocity increases is hindered by the emergence of a new phenomenon, which we call {\it solitary waves}, characterized by the slow, nearly periodic formation of large amplitude isolated fluid bumps pushed by the wind. Figure~\ref{fig:solitarywave} shows a picture of such solitary wave; the associated slopes are far above the limit for FS-SS surface reconstruction and the FS-SS technique is no longer applicable.

Solitary waves are typically 5~mm high, 2-3 cm wide in the $x$ direction, with a steep rear and a weaker slope at the front. Similar rear-front asymmetry is also found in waves in viscous liquids with strong lateral confinement~\cite{Meignin_2001}. This third regime, which overlaps with the regular wave regime, apparently corresponds to a distinct physical process: for identical viscosity and wind velocity, the typical distance between solitary waves is at least 4 times larger than the wavelength of regular waves, and their propagation velocity twice smaller. Their finite amplitude even very close to the onset suggests a subcritical mechanism for their formation. This solitary wave regime is not characterized in detail in the following, and we focus on the wrinkles and regular waves regimes.

\subsection{The wrinkles regime}
\label{sec:wr}

We first focus on the shape and amplitude of the surface deformations in the wrinkle regime. In order to characterize the typical dimensions of the wrinkles, we compute the two-point correlation of the the surface height, $C({\bf r}) = \langle \zeta({\bf x},t) \zeta({\bf x} + {\bf r},t) \rangle / \langle \zeta({\bf x},t)^2 \rangle$, where $\langle \cdot \rangle$ is a temporal and spatial average. We define the correlation length $\Lambda_i$ in the direction ${\bf e}_i$ ($i=x,y$) as 6 times the first value of $r_i$ satisfying $C(r_i) = 1/2$. This definition is chosen such that $\Lambda_i$ coincides with the wavelength for a monochromatic wave propagating in the direction ${\bf e}_i$. Although wrinkles do not have a well-defined wavelength, $\Lambda_x$ and $\Lambda_y$ provide estimates for their characteristic dimensions in the streamwise and spanwise directions.

\begin{figure}[htbp]
	\begin{center}
		\includegraphics[width=10cm]{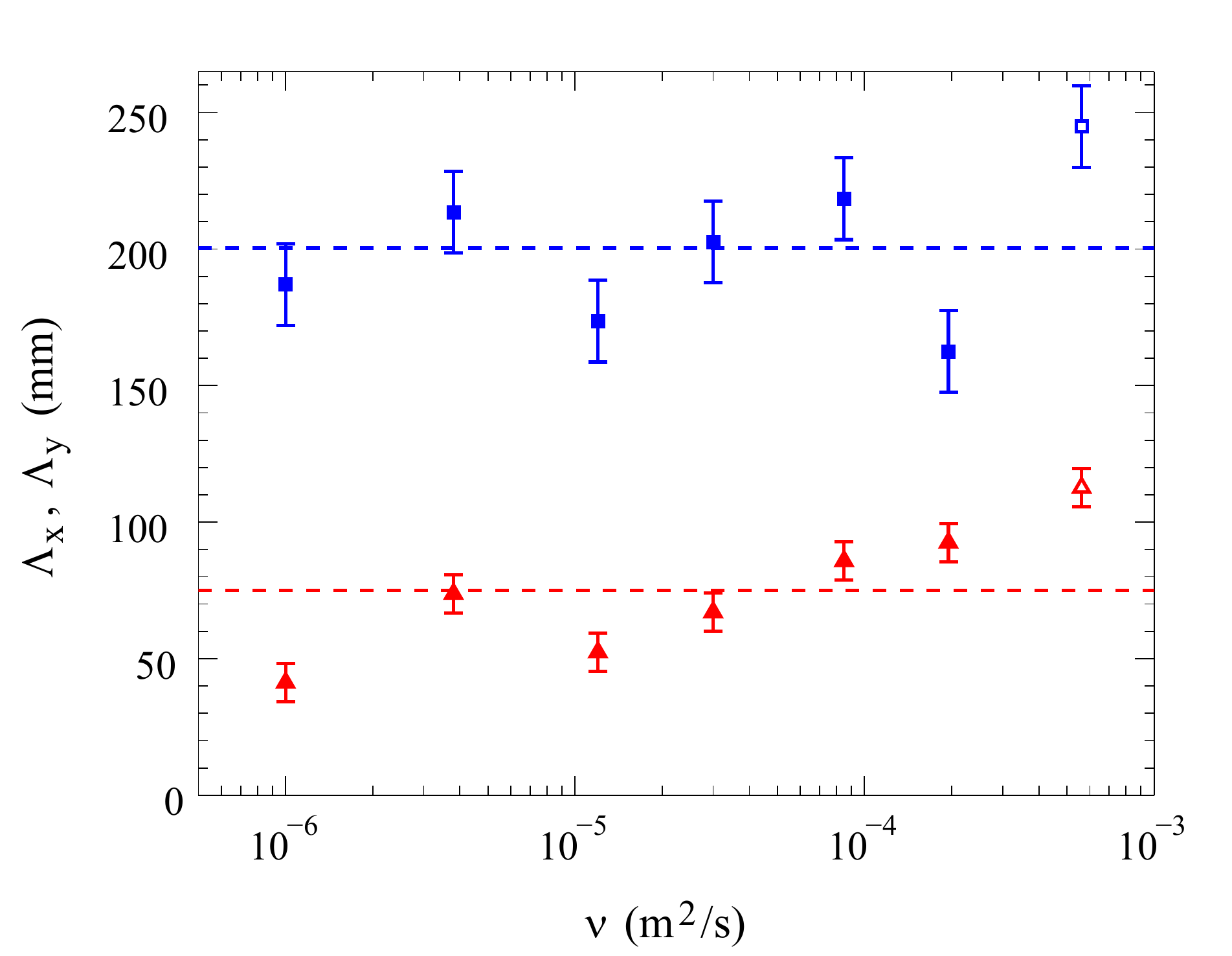}
		\caption{Streamwise and spanwise correlation lengths, $\Lambda_x$ ($\color{blue}\blacksquare$) and $\Lambda_y$ ($\color{red}\blacktriangle$), in the wrinkles regime as a function of the liquid viscosity. For all but the highest viscosity, the correlation lengths are taken at a wind velocity $U_a = 3.2 \pm 0.2$~\mps (filled symbols). For the highest viscosity, $\zeta_{\rm rms}$ is dominated by the background noise $\zeta_{\rm noise}$ at this wind velocity, so the data at the smallest available velocity in the wrinkles regime ($U_a = 4.5$~\mps) is used instead (open symbols). The dotted lines correspond to the average values $\Lambda_x \simeq$~200~mm and $\Lambda_y \simeq$~75~mm.}
	\label{fig:correlation_lengths_vs_nu}
	\end{center}
\end{figure}

The correlation lengths, plotted in Fig.~\ref{fig:correlation_lengths_vs_nu}, are remarkably constant over the whole range of liquid viscosity, with average values $\Lambda_x \simeq$~200~mm and $\Lambda_y \simeq$~75~mm. These values are obtained for a wind velocity $U_a \simeq 3.2$~\mps, but other velocities in the wrinkle regime yield similar results. This suggests that the characteristic size of the wrinkles is not a property of the flow in the liquid, but is rather geometrically constrained by the thickness of the turbulent boundary layer in the air. This thickness, as measured by $\delta_{0.99}$ (the distance at which the mean velocity is $0.99 U_a$, see Fig.~\ref{fig:system}), is of order of 15-20~mm at small fetch~\cite{Paquier_2015}. Such aspect ratio $\Lambda_x / \delta_{0.99} \simeq 10$ is indeed typical of large-scale streamwise vortices in turbulent boundary layers~\cite{Jimenez_2008}, confirming that the wrinkles can be viewed as the traces of the stress fluctuations traveling in the air flow.

We now turn to the influence of viscosity and wind velocity on the amplitude of the wrinkles. In the following, we make use of the friction velocity $u^*$, which is a more relevant parameter close to the interface than the mean wind velocity $U_a$. The friction velocity is defined as $u^* = \sqrt{\sigma/\rho_a}$, with $\sigma$ the shear stress at the surface, which can be inferred from the surface drift velocity.  Considering the air flow as a canonical turbulent boundary layer over a no-slip flat wall,  at least up to the wave generation threshold, the wind profile can be locally described with a classical logarithmic law, leading to the relationship (e.g., Ref.~\cite{Cousteix_2007})
\begin{equation}
U_a = \left[\dfrac{1}{\kappa} \log Re_{\tau} + C^+ \right] u^*,
\label{eq:Ua_vs_uf_log}
\end{equation}
with $Re_{\tau} = H u^* / 2 \nu_a$ the half-height channel turbulent Reynolds number ($\nu_a$ is the kinematic viscosity of air), $\kappa=0.4$ the von K\'arm\'an constant, and $C^+ = 5$. For the range $Re_{\tau} = 250-1500$ explored here, this law yields approximately $u^* \simeq 0.05 U_a$ to within 20\%, in agreement with the hot-wire and PIV measurements reported in Ref.~\cite{Paquier_2015}.

\begin{figure}[htb]
	\begin{center}
		\includegraphics[width=0.9\columnwidth]{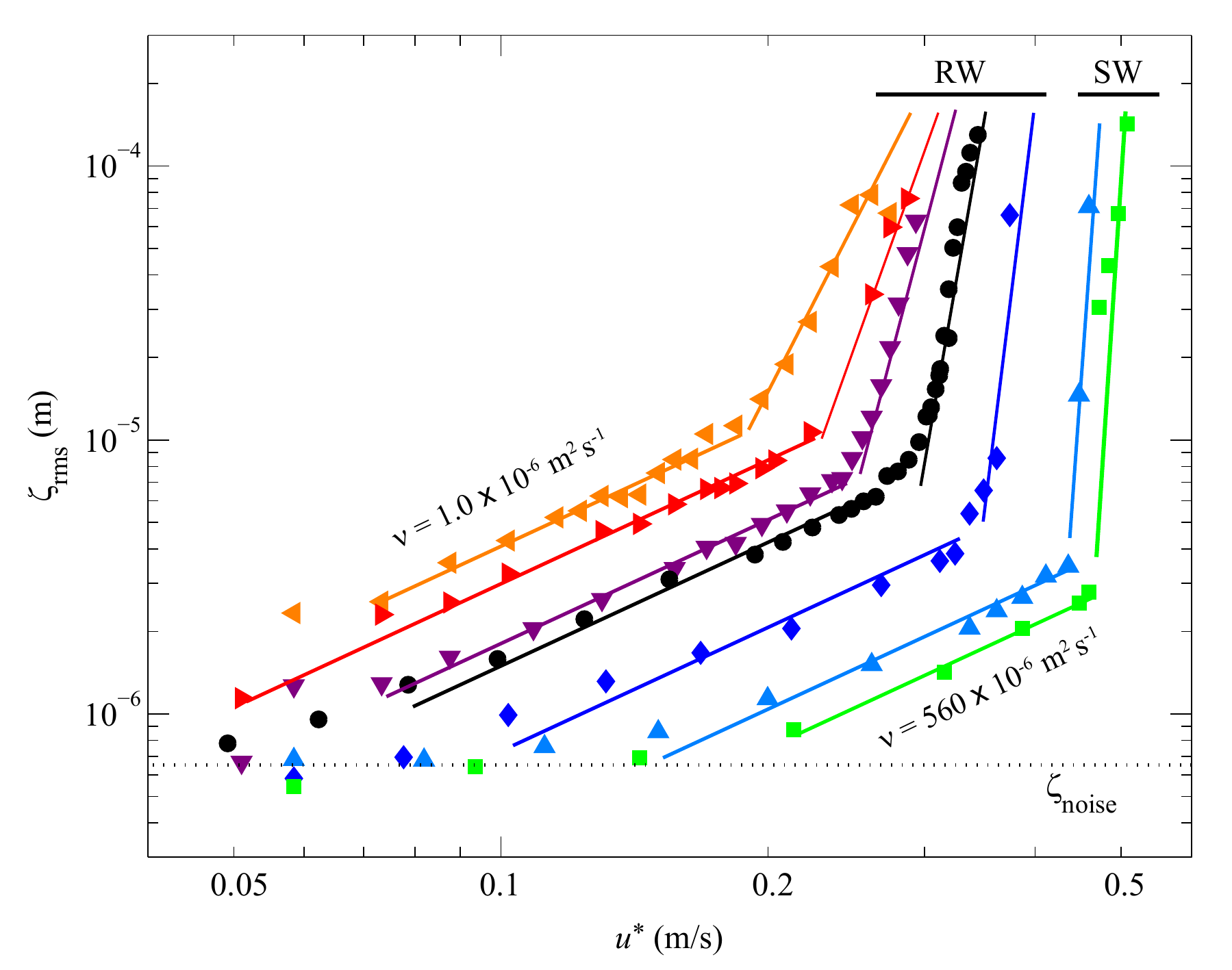}
		\caption{Amplitude of the surface deformation $\zeta_{\rm rms}$ as a function the friction velocity $u^*$ for different liquid viscosity (from top to bottom, $\nu =$~1.0, 3.9, 12, 30, 85, 195 and 560~$\times 10^{-6}$~m$^2$~s$^{-1}$).  Below the threshold (wrinkles regime), the continuous lines are power laws $\zeta_{\rm rms} \propto u^{*3/2}$. Above the threshold, data correspond to the regular wave regime (RW) at small viscosity, and to the solitary wave regime (SW) at high viscosity;  the continuous lines are only guides to the eye. The horizontal dotted line gives the background noise $\zeta_{\rm noise} = 0.65$~$\mu$m.}
		\label{fig:Rms_fits_uf}
	\end{center}
\end{figure}

The surface deformation amplitude $\zeta_{\rm rms}$ is plotted as a function of the friction velocity $u^*$ in Fig.~\ref{fig:Rms_fits_uf} for the full range of fluid viscosities. This amplitude is defined here in a statistical sense, as the root mean square of the surface height $\zeta_{\rm rms} = \langle \zeta^2(x,y,t) \rangle^{1/2}$, where the brackets represent both a temporal and a spatial average. The data show a clear transition between the wrinkle regime at low velocity and the sharply increasing wave regime at larger velocity (this wave regime includes here both the regular waves up to $\nu \simeq 100-200 \times 10^{-6}$~m$^2$~s$^{-1}$ and the solitary waves at larger $\nu$). This plot confirms that, at a given wind velocity, a larger fluid viscosity leads to weaker surface deformations.  At very large viscosity, the wrinkles amplitude becomes very weak and saturates to the measurement noise level, which we can estimate as $\zeta_{\rm noise} \simeq 0.65~\mu$m. This noise originates from slight residual vibrations in the wind-tunnel or in the acquisition setup, and from reconstruction errors in the FS-SS data processing.  At low viscosity, the surface deformations are always greater than this background noise, even at the lowest wind velocity.

\begin{figure}[htb]
	\begin{center}
		\includegraphics[width=10cm]{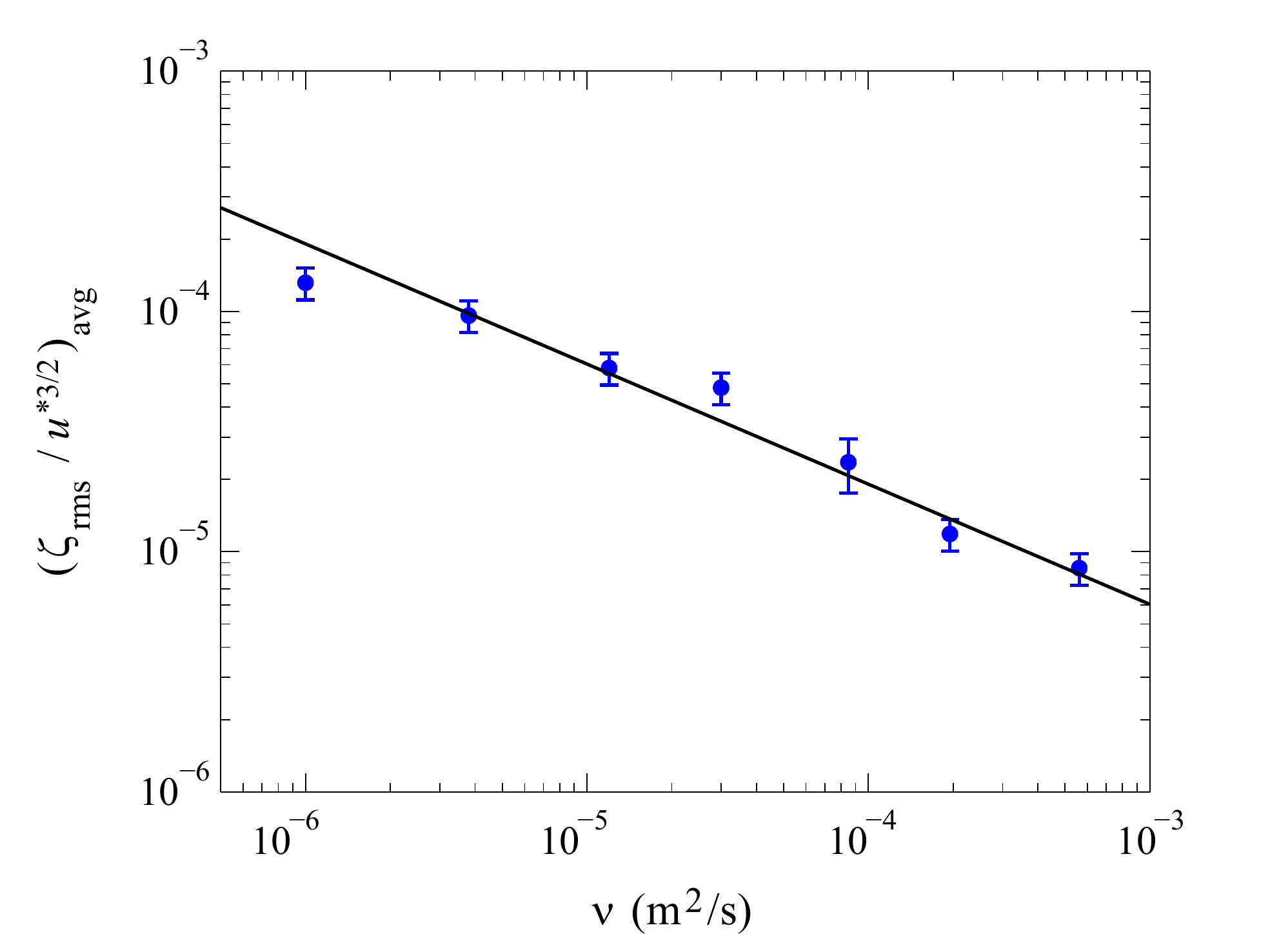}
		\caption{Ratio $\zeta_{\rm rms}/u^{*3/2}$ in the wrinkles regime as a function of the kinematic viscosity. The line shows the fit $\alpha \nu^{-1/2}$ with $\alpha = 1.9 \times 10^{-7}$ SI.}
		\label{fig:Coeff_power_fit_uf}
	\end{center}
\end{figure}

For friction velocities between the noise-limited lower bound and the wave threshold, the amplitude of the wrinkles grows approximately as a power law, $\zeta_{\rm rms} \propto u^{*m}$, with $m = 1.5 \pm 0.15$ for all viscosities.
Note that plotting the same data versus wind velocity $U_a$ leads to a slightly shallower power law, $\zeta_{\rm rms} \propto U_a^{m'}$ with $m' = 1.1 \pm 0.15$, consistent with the approximately linear evolution found in Paquier {\it et al.}~\cite{Paquier_2015}; this discrepancy between the exponents $m$ and $m'$ can be ascribed to the limited range of velocity and to the logarithmic correction in the relation (\ref{eq:Ua_vs_uf_log}) between $u^*$ and $U_a$.

In order to evaluate the dependence with respect to $\nu$ of the wrinkles amplitude, we finally plot in Fig.~\ref{fig:Coeff_power_fit_uf} the ratio $\zeta_{\rm rms} / u^{*1.5}$ averaged over the range of $u^*$ for which the power law is observed. This ratio is found to decrease approximately as $\nu^{-0.5 \pm 0.05}$, suggesting the general scaling law for the wrinkles amplitude
\begin{equation}
\zeta_{\rm rms} \propto \nu^{-1/2} u^{*3/2}.
\label{eq:zrma_over_ufpower15}
\end{equation}
This scaling holds over nearly the whole viscosity range, except for the lowest viscosity (water) which shows wrinkles of slightly smaller amplitude. A phenomenological model for this scaling is proposed in Sec.~\ref{sec:model}.

\subsection{The regular wave regime}
\label{thresholds}

For friction velocity larger than a viscosity-dependent threshold $u^*_c$, the system enters the wave regime and transverse waves start to emerge (right column of Fig.~\ref{fig:FSSS_different_visc}). Figure~\ref{fig:Rms_fits_uf} shows a much sharper increase of $\zeta_{\rm rms}$ with $u^*$ in the wave regime than in the wrinkle regime, and this sharp increase is even more pronounced as viscosity is increased. Interestingly, the smooth transition observed at low viscosity may explain some of the discrepancies found on the thresholds reported for experiments performed in water, which we discuss in Sec.~\ref{sec:lit}.

\begin{figure}[htb]
	\begin{center}
		\includegraphics[width=10cm]{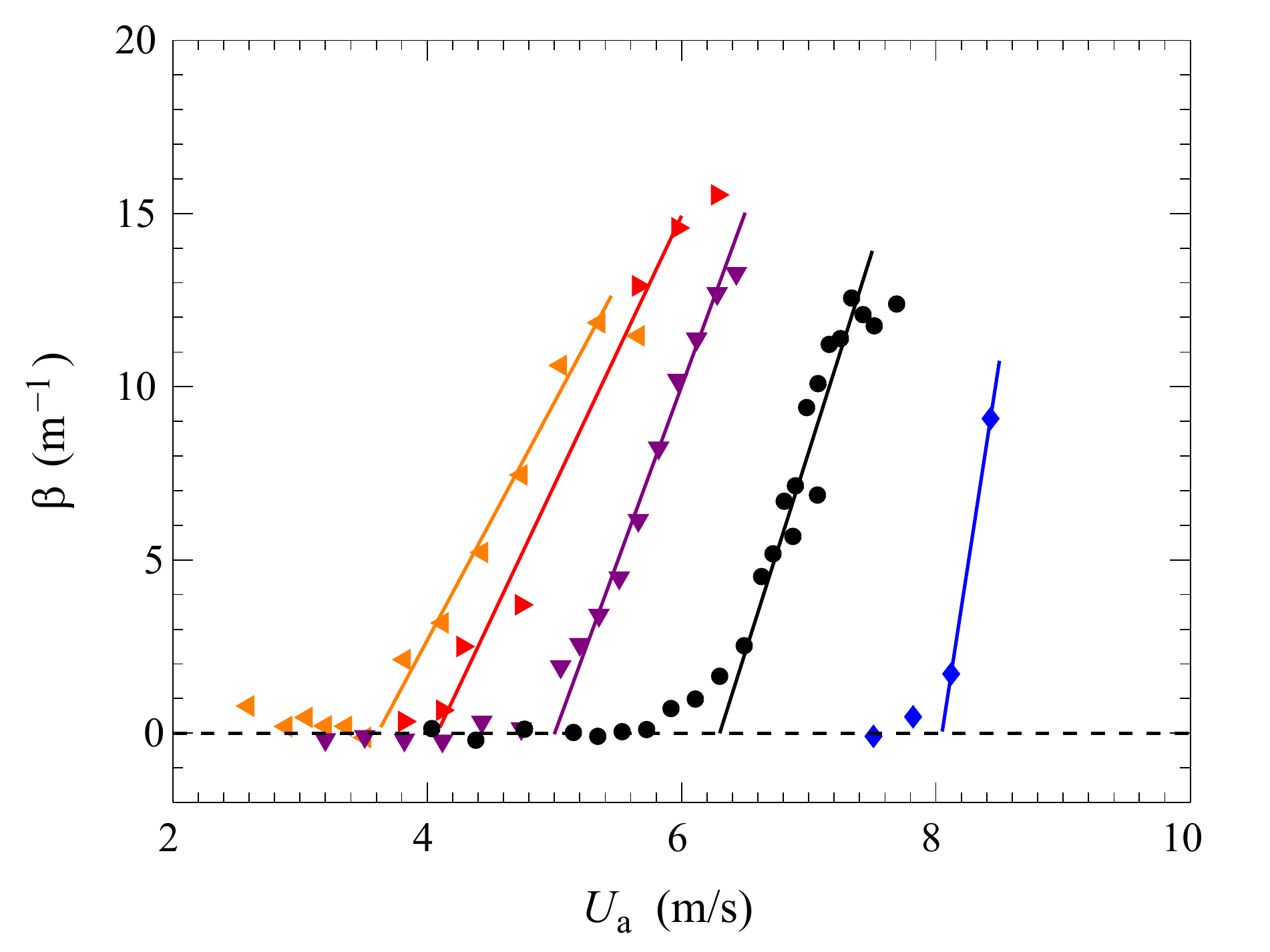}
		\caption{Spatial growth rate $\beta$ of the regular waves as a function of the wind velocity $U_a$ for different viscosities (same symbols as in Fig.~\ref{fig:Rms_fits_uf}). The crossing of the linear fits with $\beta=0$ defines the critical wind velocity $U_c$.}
	\label{fig:beta_vs_Ua}
	\end{center}
\end{figure}

In order to accurately define the threshold velocity, we compute the spatial growth rate $\beta$ from the exponential growth of the squared wave amplitude, $\zeta^2_{\rm rms}(x) \propto \exp(\beta x)$, at short fetch~\cite{Paquier_2015}. The growth rate is plotted in Fig.~\ref{fig:beta_vs_Ua} as function of $U_a$ for different viscosities. We obtain $\beta \simeq 0$ below the wave onset (in the wrinkle regime), and an approximately linear increase of $\beta$ with $U_a$ above the threshold. As for the wave amplitude itself, the transition to positive growth rates becomes more abrupt as the viscosity is increased: the rate of increase of $\beta$ with $U_a$ is four times larger at the largest viscosity than in water. The threshold velocity $U_c$ can be finally defined as the crossing of the linear fit with $\beta  = 0$.  Note that for the two highest viscosities ($\nu > 200 \times 10^{-6}$~m$^2$~s$^{-1}$), this method cannot be used due to the presence of solitary waves of large amplitude which prevents FS-SS measurements. In these cases, the velocity threshold is simply estimated by the intersection of the power-law fits of $\zeta_{\rm rms}$ in the wrinkles and wave regimes. To check the determination of $U_c$ from FS-SS data, we also measured the threshold velocity using the reflection of a tilted laser beam on the surface. Indeed, the vertical displacement of the laser dot on a screen can be related to the local slope $S = \partial \zeta / \partial x$ of the surface, and the rapid change of slope offers a good evaluation of the threshold velocity. Measurements of the thresholds by both methods yield similar results.

\begin{figure}[htb]
	\begin{center}
		\includegraphics[width=13cm]{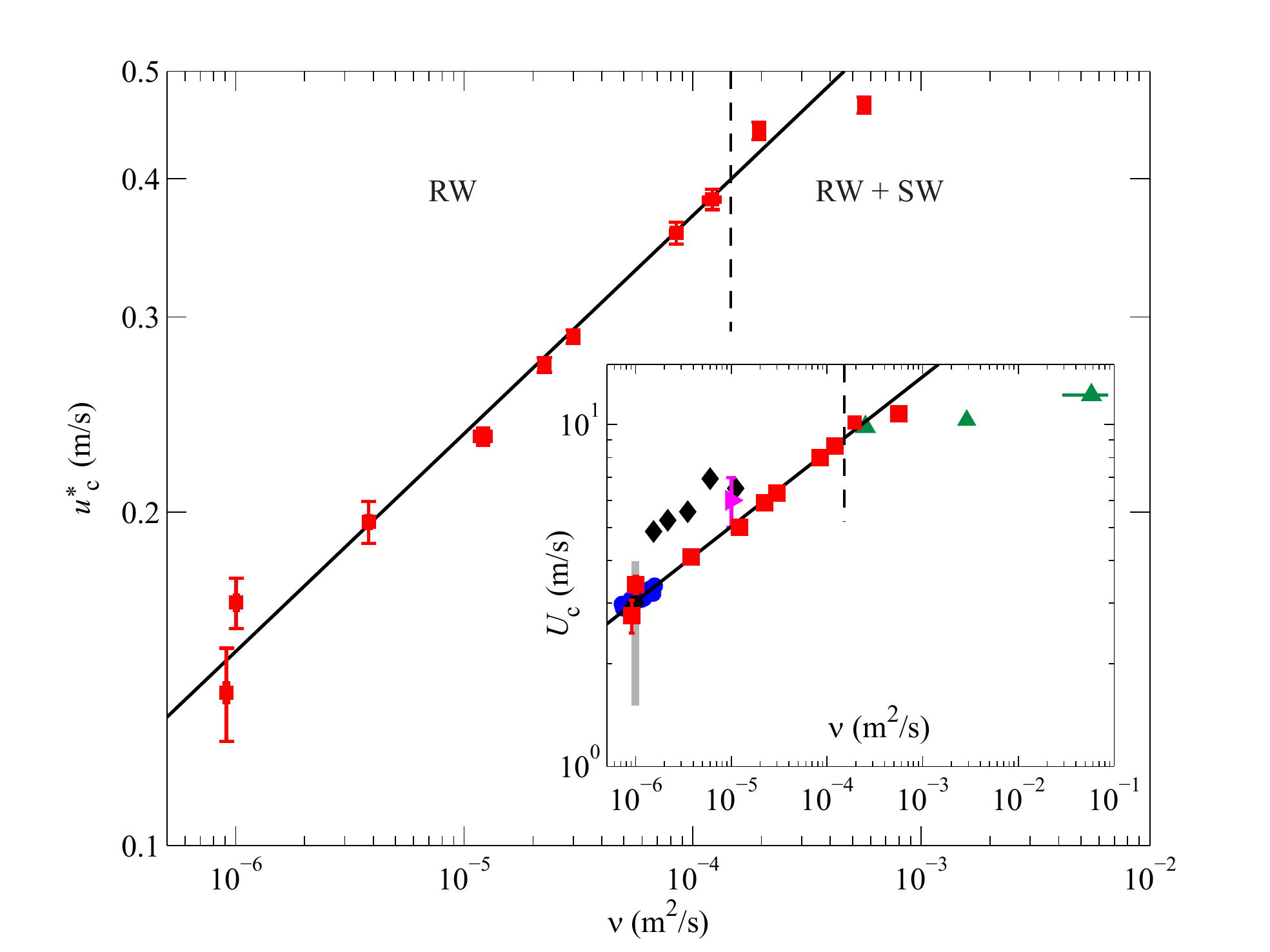}
		\caption{Critical friction velocity $u_c^*$ at wave onset as a function of the kinematic viscosity of the liquid. The continuous line shows the fit $u_c^* = a \nu^{0.20}$, with $a = 2.3$ SI, in the regular wave (RW) regime; SW denotes the solitary wave regime.
Inset: $\color{red}\blacksquare$, same data as in the main figure, expressed in terms of the critical mean air velocity $U_c$, compared with data taken from the literature (see Tab.~\ref{tab:thresholds_literature}): $\blacklozenge$, Keulegan~\cite{Keulegan_1951}; $\color{blue}\bullet$, Kahma and Donelan~\cite{Kahma_1988}; $\color{magenta}\blacktriangleright$, Gottifredi and Jameson~\cite{Gottifredi_1970};  $\color{green}\blacktriangle$, Francis~\cite{Francis_1956} (an arbitrary error bar of $\pm50$\% of the viscosity has been added because of the large uncertainty due the hygroscopic nature of his most viscous liquid).  The vertical gray bar represents the range of thresholds from the literature obtained obtained in water laboratory experiments. The continuous line is a power fit $U_c = A \nu^{0.22}$, with $A = 62.6$~SI. Note that error bars are often smaller than the markers.}
	\label{fig:thresholds_uf}
	\end{center}
\end{figure}

The evolution of the critical friction velocity $u^*_c$ (deduced from $U_c$ using Eq.~(\ref{eq:Ua_vs_uf_log})) with liquid viscosity is shown in Fig.~\ref{fig:thresholds_uf}, confirming that a more viscous fluid requires a stronger wind to trigger wave generation. A well-defined power law is found,
\begin{equation}
u_c^* \propto \nu^{0.20 \pm 0.01},
\label{eq:ucvsnu}
\end{equation}
at least before the solitary wave transition (i.e., for $\nu < 100-200 \times 10^{-6}$~m$^2$~s$^{-1}$).  The critical velocity for the highest viscosity (which is not included in the fit) suggests a slightly shallower increase with $\nu$ for solitary waves. The validity of the power law (\ref{eq:ucvsnu}) down to $\nu = 10^{-6}$~m$^2$~s$^{-1}$ (water) is remarkable: Indeed, for such low viscosity, we may expect different physical mechanisms for the onset of wave growth --- the surface drift is larger, the flow in the liquid can be unstable, and waves propagate over a longer distance before being damped, allowing for reflections at the end of the tank.

\begin{figure}[htb]
	\begin{center}
		\includegraphics[width=\columnwidth]{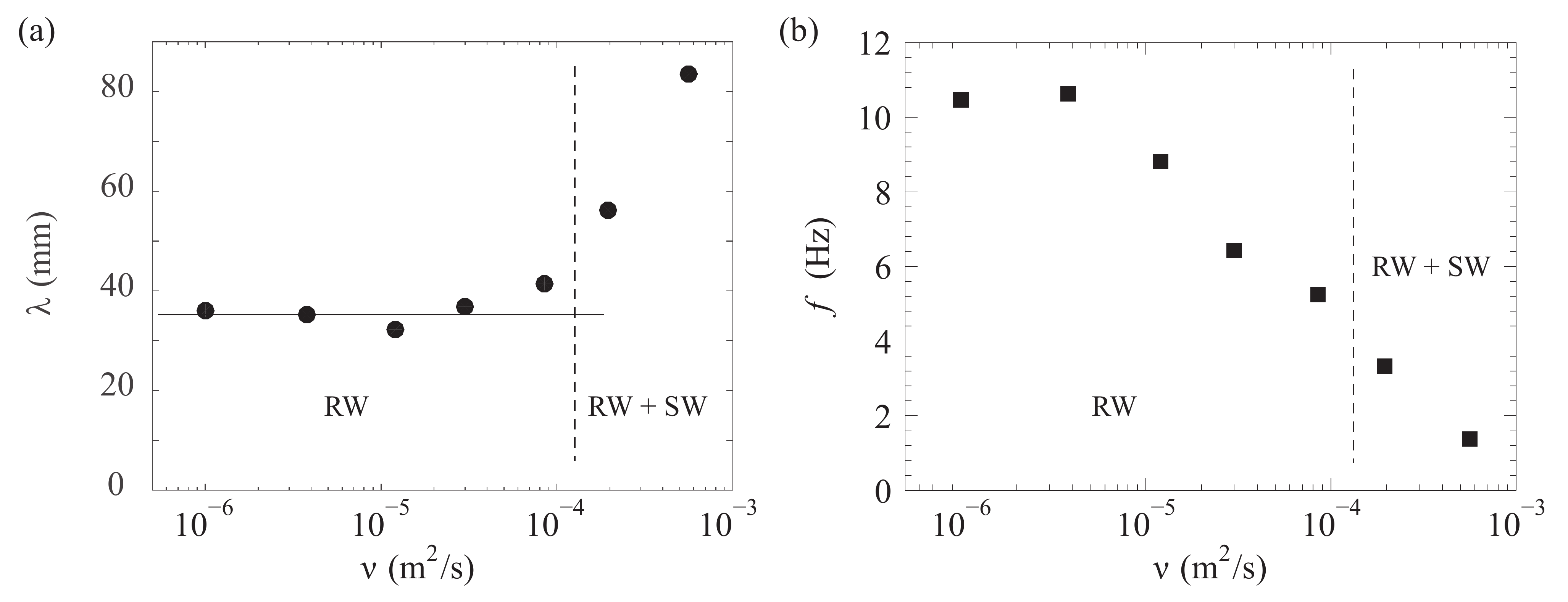}		
		\caption{Critical wavelength $\lambda$ (a) and frequency $f$ (b) measured at wave onset as a function of the kinematic viscosity of the liquid.}
	\label{fig:lambda_and_f_vs_nu}
	\end{center}
\end{figure}

We finally show in Fig.~\ref{fig:lambda_and_f_vs_nu} the influence of the liquid viscosity on the wavelength and frequency of the first waves generated at the critical wind velocity. The wavelength is computed from two-point correlation (see Sec.~\ref{sec:wr}), and a similar procedure (two-time correlation at a fixed point) is applied to compute the frequency. The critical wavelength is almost independent of the viscosity, $\lambda_c \simeq 35$~mm, over the whole range for which regular waves are observed, up to $\nu \simeq 10^{-4}$~m$^2$~s$^{-1}$. The strong increase of the wavelength at larger viscosity corresponds to the onset of solitary waves. On the other hand, the critical frequency is found to continuously decrease, with no evidence of transition between the regular and solitary wave regimes.

\subsection{Comparison with the literature}
\label{sec:lit}

\begin{table}
\begin{center}
\caption{Threshold air velocities for wave generation from the literature. Concerning experiments carried out in water, when the viscosity of the water is available in the reference paper, the value is given as in the table. If only the water temperature is mentioned, its viscosity is taken as the tabulated value of pure water at this temperature. Otherwise, when neither the viscosity nor the temperature of the water during their wind waves experiments is indicated by the authors, the kinematic viscosity of the water is taken to be $1.0 \times 10^{-6}$~m$^2$~s$^{-1}$.}
\begin{tabular}{p{0.6cm}|p{3.6cm}|p{2.2cm}|p{1.7cm}|p{1.7cm}|p{2.5cm}}
\hline \hline
 & Reference & Liquid & Kinematic viscosity & Threshold velocity & Notes\\
& & & $\nu$ ($\times 10^{-6}$ m$^2$~s$^{-1}$) & $U_c$ (m~s$^{-1}$)\\
\hline
1 & Roll, 1951	  	                 	 & \multirow{4}{2cm}{water (outdoors)}& \multirow{4}{*}{$\sim 1$}	& 0.4 		& cited by Ref.~\cite{Kahma_1988} \\
2 & Russell, 1844 \cite{Russell_1844}  &      			& 	 	& 0.85 		    & cited by Ref.~\cite{Kahma_1988} \\
3 & Jeffreys, 1925 \cite{Jeffreys_1925}&   	        & 		& $1.0-1.2$ 	&  \\
4 & Van Dorn, 1953  	                 & 						& 	 	& 2			      & cited by Ref.~\cite{Kahma_1988} \\ 
\hline
5 & Kahma and Donelan, 1988 \cite{Kahma_1988}    	& \multirow{4}{2cm}{water (laboratory)}	& $0.72-1.58$	& $2.9-3.4$		&  \\
6 & Donelan and Plant, 2009 \cite{Donelan_2009}	  &     		& 0.81		& $\sim 1.5$ &  \\
7 & Keulegan, 1951 \cite{Keulegan_1951}           &    			& 0.92		& 2.9		     &  \\
8 & Donelan and Plant, 2009 \cite{Donelan_2009}		& 		    & 1.00		& $\sim 1.7$ &  \\
\hline
9 & Wu, 1978 \cite{Wu_1978}             	&	\multirow{5}{2cm}{water (laboratory)}		&\multirow{5}{*}{$\sim 1$} & 1.6		&  \\
10& Hidy and Plate, 1966 \cite{Hidy_1966}	&       		&		& 3			&  \\
11& Francis, 1951	\cite{Francis_1951}		  &           &		& $3.1-3.2$	&  \\ 
12& Gottifredi and Jameson, 1970 \cite{Gottifredi_1970}& 		&		& 3.5 		& $U_c$ based on visual wave growth\\
13& Lorenz {\it et al.}, 2005 \cite{Lorenz_2005}  &			&		& 4-6		&  \\ 
\hline
14& \multirow{5}{*}{Keulegan, 1951 \cite{Keulegan_1951}} & \multirow{5}{2cm}{sugar solutions of different concentrations}  & 1.54 & 4.9	& \multirow{5}{2.3cm}{$U_c$ is the average of values taken at different fetches} \\
15&  					&    & 2.17	& 5.2	&  \\
16& 					& 	 & 3.51	& 5.6	&  \\
17&						&    & 6.00	& 6.9	&  \\
18&						& 	 & 11.1	& 6.50 	&  \\
\hline
19&Gottifredi \& Jameson, 1970 \cite{Gottifredi_1970}  & glycerol-water solution & 10.0	& $\sim 6$ & $U_c$ based on visual wave growth\\
\hline
20& \multirow{3}{*}{Francis, 1956 \cite{Francis_1956}}	 & oil		& 250	& 9.84		&  \\ 
21& 					& oil		& 2~900	& 10.30		&  \\ 
22&						& syrup 	& 58~000	& 12.20		& uncertainty on $\nu$ \\ 
\hline \hline
\end{tabular}
\label{tab:thresholds_literature}
\end{center}
\end{table}

We now compare the critical velocities for wave onset obtained in the present study to the ones reported in the literature, which we summarize in Tab.~\ref{tab:thresholds_literature}. Since the friction velocity is often not reported by the authors, we also plot in the inset of Fig.~\ref{fig:thresholds_uf} our data in terms of the critical mean wind velocity $U_c$.

Despite the variations between the results of the different authors, Fig.~\ref{fig:thresholds_uf} shows a good overall agreement with the data from the literature.  In Kahma and Donelan~\cite{Kahma_1988} (blue dots), the viscosity was varied using water at temperature ranging from 4 to 35$^{\circ}$C. The highest viscosities are reported by Francis~\cite{Francis_1956} (green triangles), using viscous oils and a syrup 58~000 more viscous than water. Interestingly, his data points increase with a shallower slope than our $\nu^{0.2}$ law, which is compatible with our last data point at $\nu = 560 \times 10^{-6}$~m$^2$~s$^{-1}$. This suggests that the waves in the experiment of Francis are in the solitary wave regime~\cite{Francis_1954}. Keulegan~\cite{Keulegan_1951} (black diamonds) also offers some results for intermediate viscosities. Unlike what we observe experimentally, he witnesses a critical velocity decreasing with fetch along his 20~m long tank. He thus takes $U_c$ as the average of the critical velocity at three different fetches, which may explain the fact that his results are about 40\% above ours. Despite being focused on the growth of mechanically-generated waves amplified by wind, Gottifredi and Jameson's study~\cite{Gottifredi_1970} over water or aqueous glycerol solutions also mentions the critical wind velocity in the absence of artificial waves (magenta triangle). This wind velocity is however based on the visual observation of the wave growth, which may explain onset velocities above ours, as weak wave growths may not have been visible.

Finally, `natural' wind waves (waves generated by wind in outdoors conditions, for example on a lake or at sea) present a critical velocity that is usually lower than for wind waves in laboratory (including our experiments). This may be interpreted by the fact that natural conditions can be quite different from the ones in laboratory: unbounded non stationary airflow, turbulent flow in the liquid, presence of unsteady currents under the surface, etc.  Thresholds over water in outdoors conditions are listed in Tab.~\ref{tab:thresholds_literature}, and represented as a vertical gray bar in Fig.~\ref{fig:thresholds_uf}.

\section{A model for the scaling of the wrinkles amplitude}
\label{sec:model}

We finally propose here a model for the observed scaling of the wrinkles amplitude with friction velocity and liquid viscosity, $\zeta_{\rm rms} \propto \nu^{-1/2} u^{*3/2}$ (\ref{eq:zrma_over_ufpower15}).  The wrinkles being the response of the liquid to the stress fluctuations at the surface, we expect a relation between the wrinkles amplitude rms and the stress rms. Both normal (pressure) and shear stress fluctuations are related to the downward flux of momentum from the air to the surface, and are expected to scale as $\rho_a u^{*2}$, with $\rho_a$ the air density. This scaling is confirmed by DNS of turbulent channel flow with no-slip flat walls~\cite{Hu_2006,Jimenez_2008}: for the typical turbulent Reynolds numbers considered here, $Re_\tau \simeq 160-1600$, the pressure rms at the surface is $p_{\rm rms} \simeq 2 \rho_a u^{*2}$, with a weak logarithmic correction in $Re_\tau$ which can be neglected here.

We first note that the wrinkles cannot correspond to a simple hydrostatic response to the pressure fluctuations: such response would yield $\zeta_{\rm rms} \simeq p_{\rm rms} / \rho g$ (with $\rho$ the fluid density), with no dependence on the liquid viscosity, which contradicts our observations. Inversely, wrinkles cannot either correspond to a purely creeping flow forced by the stress fluctuations: if only $u^*$, $\nu$ and the size of the pressure fluctuations were relevant parameters, dimensional analysis would require $\zeta_{\rm rms}$ to be a function of $\nu/u^*$, which again contradicts our observations. A minimum model for the wrinkles amplitude should therefore contain both gravity and viscosity effects.

In a statistically steady state, we can assume a balance between the energy loss in the wrinkles by viscous diffusion and the vertical energy flux from the turbulent fluctuations. Since the size of the wrinkles is significantly larger than the capillary length $\lambda_c \simeq 14$~mm (one has $\Lambda_x \simeq 3 \Lambda_y \simeq 15 \lambda_c$, see Fig.~\ref{fig:correlation_lengths_vs_nu}), we can neglect the surface tension and simply write the energy (potential and kinetic) per unit surface of the wrinkles as $e_w \simeq \rho g \zeta_{\rm rms}^2$. Its rate of change by viscous diffusion can be estimated as $\nu e_w/\Lambda_y^2$, assuming that the viscous time scale is governed by the smallest extent of the wrinkle, in the spanwise direction (the thickness of the boundary layer $\delta_{0.99}$, which governs the size of the wrinkles, could be used instead; see Sec.~\ref{sec:wr}). The vertical energy flux from the turbulent boundary layer to the liquid surface can be estimated as the work per unit time of the stress fluctuations, of order $\rho_a u^{*2}$, with characteristic vertical velocity fluctuations above the viscous sublayer of order $u^*$. Balancing this energy flux $\rho_a u^{*3}$ with the energy loss $\nu e_w/\Lambda_y^2$ yields
\begin{equation}
\zeta_{\rm rms} \simeq K \left(\frac{\rho_a}{\rho} \right)^{1/2} \frac{u^{*3/2} \Lambda_y}{\sqrt{g \nu}},
\label{eq:model}
\end{equation}
where $K$ is a non-dimensional constant, which is consistent with the observed scaling (\ref{eq:zrma_over_ufpower15}).

\begin{figure}[htb]
	\begin{center}
		\includegraphics[width=10cm]{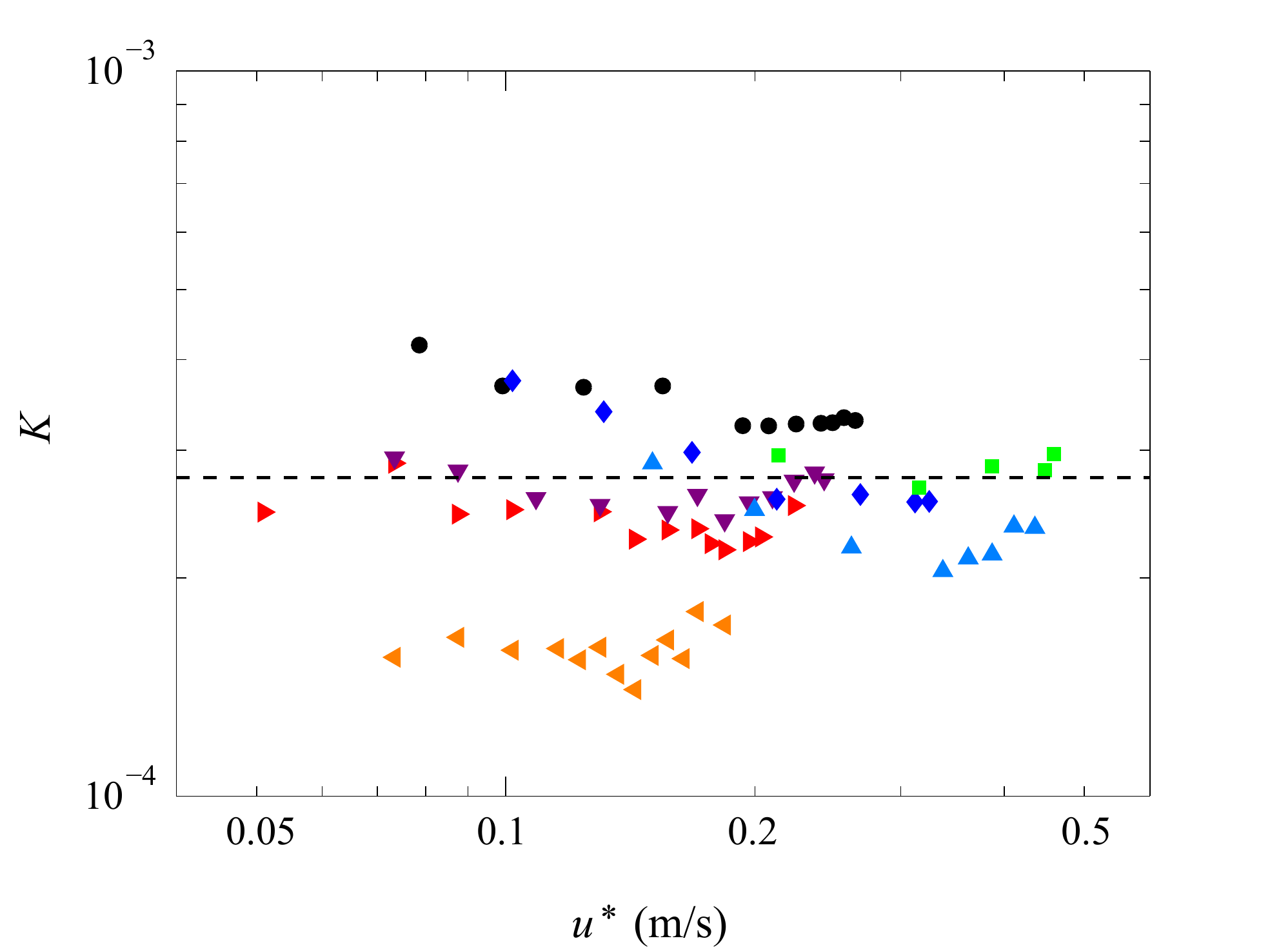}
		\caption{Coefficient $K$ in Eq.~(\ref{eq:model}), representing the normalized wrinkles amplitude, as a function of the friction velocity for different liquid viscosities (same symbols as in Fig.~\ref{fig:Rms_fits_uf}). The dashed line corresponds to the average $K \simeq 3 \times 10^{-4}$.}
	\label{fig:K_vs_uf}
	\end{center}
\end{figure}

Figure~\ref{fig:K_vs_uf}, where the numerical factor $K$ is plotted for the range of velocities for which the power law $u^{*3/2}$ holds, shows a fairly good collapse of the data around the average value $K \simeq (3 \pm 1) \times 10^{-4}$ . The main discrepancy to this law corresponds to the experiments performed with water, which again show significantly smaller wrinkles amplitude.
The reason for this discrepancy might be that, at low viscosity, a significant amount of the turbulent energy flux is transferred to the surface drift current, resulting in weaker surface deformations \cite{Grare_2013}. Flow instabilities induced by the turbulent stress fluctuations  may also yield extra dissipation in the liquid, and hence weaker surface deformations.

Interestingly, a scaling similar to Eq.~(\ref{eq:model}) can be obtained by modifying the theory of Phillips~\cite{Phillips_1957} to include the effects of the liquid viscosity. Assuming an inviscid liquid, Phillips derives the amplitude of  the surface deformations generated by a statistically steady turbulent airflow applied over the
surface during a time $t$,
\begin{equation}
\zeta_{\rm rms}^2 \simeq \frac{p_{\rm rms}^2 t}{2 \sqrt{2} \rho^2 V_c g},
\label{eq:phil}
\end{equation}
where $V_c$ is the characteristic convection velocity of the pressure fluctuations. We can include qualitatively the effect of the liquid viscosity in his analysis, by assuming that the temporal growth in Eq.~(\ref{eq:phil}) saturates on a viscous time scale $t \simeq \Lambda_y^2 / \nu$. With this assumption, and taking $V_c \propto u^*$, Eq.~(\ref{eq:phil}) yields a scaling also consistent with the observed behavior $\zeta_{\rm rms} \propto \nu^{-1/2} u^{*3/2}$.

\section{Conclusion}
\label{sec:conclusion}

In this paper, we study experimentally the influence of the liquid viscosity on the early stages of wind wave generation.  Most of the results of Paquier {\it et al.}~\cite{Paquier_2015} obtained for a single viscosity, in particular the existence of two regimes of surface deformation, the wrinkles regime and the wave regime, are extended here over a wide range of viscosities.

For all viscosities, the surface below the wave onset is populated  by disorganized elongated wrinkles whose characteristic amplitude scales as $\nu^{-1/2} u^{*3/2}$, which correspond to the trace of the turbulent stress fluctuations at the surface. A simple model based on an energy balance between the turbulent energy flux and the viscous dissipation in the liquid is proposed to account for this scaling.

The critical velocity for wave onset is found to slowly increase as $u_c^* \sim \nu^{0.2}$ over nearly two decades of viscosity, in accordance with data from the literature. Whereas the transition from wrinkles to waves is abrupt at large viscosity, it is much smoother at low viscosity, in particular for water. Interestingly, this smooth transition may explain, at least in part, the large scatter in the critical wave onset velocity reported in the literature for water waves.

Finally, a new regime is found in highly viscous liquids ($\nu > 100-200 \times 10^{-6}$~m$^2$~s$^{-1}$):  At sufficiently high wind velocity, in addition to the regular wave regime, a solitary wave regime appears, characterized by slow, nearly periodic formation of large-amplitude localized fluid bumps.  This highly nonlinear regime could not be characterized  with the present FS-SS measurements, limited to small wave slope and curvature, and warrants further investigation.

\acknowledgments

We are grateful to H. Branger, C. Clanet, F. Charru, A. Hector, and P. Spelt for fruitful discussions. We acknowledge J. Amarni, A. Aubertin, L. Auffray, and R. Pidoux for experimental help. This work is supported by RTRA ``Triangle de la Physique''. F.M. acknowledges the Institut Universitaire de France.


\end{document}